\documentclass[aip,apl,twocolumn,10pt]{revtex4-1}
\usepackage{units}
\usepackage{amsmath}
\usepackage{amssymb}
\usepackage{graphicx}
\usepackage{textcomp}

\begin{document}

\title{\Large{Direct measurement of second-order coupling in a waveguide lattice}}

\author{Robert Keil}
\email{robert.keil@uibk.ac.at}
\affiliation{Institut f\"ur Experimentalphysik, Universit\"at Innsbruck, Technikerstrasse 25, 6020 Innsbruck, Austria}
\author{Benedikt Pressl}
\affiliation{Institut f\"ur Experimentalphysik, Universit\"at Innsbruck, Technikerstrasse 25, 6020 Innsbruck, Austria}
\author{Ren\'{e} Heilmann}
\affiliation{Institute of Applied Physics, Abbe Center of Photonics, Friedrich-Schiller-Universit\"{a}t Jena, Max-Wien-Platz 1, 07743 Jena, Germany}
\author{Markus Gr\"{a}fe}
\affiliation{Institute of Applied Physics, Abbe Center of Photonics, Friedrich-Schiller-Universit\"{a}t Jena, Max-Wien-Platz 1, 07743 Jena, Germany}
\author{Gregor Weihs}
\affiliation{Institut f\"ur Experimentalphysik, Universit\"at Innsbruck, Technikerstrasse 25, 6020 Innsbruck, Austria}
\author{Alexander Szameit}
\affiliation{Institute of Applied Physics, Abbe Center of Photonics, Friedrich-Schiller-Universit\"{a}t Jena, Max-Wien-Platz 1, 07743 Jena, Germany}



\date{\today}

%

\begin{abstract}
We measure the next-nearest-neighbour coupling in an array of coupled optical waveguides directly via an integrated eigenmode interferometer. In contrast to light propagation experiments, the technique is insensitive to nearest-neighbour dynamics. Our results show that second-order coupling in a linear configuration can be suppressed well below the level expected from the exponential decay of the guided modes.
\end{abstract}


\newcommand\beq{\begin{equation}}
\newcommand\eeq{\end{equation}}
\newcommand\bfig{\begin{figure}}
\newcommand\efig{\end{figure}}

\maketitle

\noindent 50 years past its proposition \cite{Jones:OpticalFibers}, evanescent coupling between optical waveguides has become a standard part in the optical engineer's toolbox and is now widely applied in science and industry \cite{Okamoto:Fundamentals}. Extended lattices of coupled waveguides have been used for various fundamental studies and applications, ranging from artificial graphene \cite{Peleg:ConicalDiffraction,Rechtsman:ArtificialMagField} and quantum walks \cite{Perets:RandomWalk,Peruzzo:QuantumWalkPhotonPairs,Caruso:QuantumMaze,Biggerstaff:EnhancedQuantumTransport} to mode-locking of lasers \cite{Chao:ModeLockingWaveguideArray} and quantum state preparation \cite{Solntsev:BiphotonStatesArray,Graefe:Wstates}. These systems are usually based on coupling between nearest neighbours. Coupling between more distant sites can often be neglected, due to an exponential decay of the waveguide modes \cite{Somekh:ChannelOpticalWaveguides}. Yet, there are configurations for which precise knowledge and control of such couplings become crucial. For instance, some one-dimensional arrays are designed towards an effective cancellation of nearest-neighbour coupling after certain distances, such that coupling between next-nearest-neighbours (henceforth termed \lq second-order coupling\rq) can become the dominant mechanism of transverse transport \cite{Pertsch:BlochOscillationTemp,Morandotti:BlochOscillation,Longhi:DynLoc,Szameit:Imaging}. Evidently, the distance between next-nearest neighbours in two-dimensional lattices does not need to be much larger than the one between nearest neighbours \cite{Rechtsman:ArtificialMagField,Caruso:QuantumMaze}. Therefore, second-order coupling is often quite relevant in such systems. Moreover, second-order coupling has been shown to have considerable impact in nonlinear optics, where it determines the existence of a power treshold for discrete solitons \cite{Szameit:ZigZagSoliton}, as well as on two-particle interference conditions in the quantum regime \cite{Qi:SecondOrderCouplingPhotonCorrelation}.
 
In order to obtain systematic knowledge of how and with which strength second-order coupling arises in the configurations of interest it would be desirable to measure it directly. However, its subtle influence on the propagation dynamics is often masked by the much stronger first-order coupling (or the unknown fidelity of its cancelling mechanism), such that it is quite hard to unambiguously extract the second-order coupling from light propagation experiments. It seems more promising to use the impact of second-order coupling on the eigenmodes of the system for experimental access. Indeed, second- and even third-order coupling in square and honeycomb lattices of microwave resonators have been unambiguously identified from their frequency spectra \cite{Bellec:MicrowaveTightBinding}. In optics, however, a direct measurement of waveguide eigenmodes is more challenging and requires interferometric techniques. In this work, we present such a method and apply it to measure second-order coupling in the most fundamental system where it can occur - an array of three waveguides. In particular, we investigate whether the second-order coupling between the outer waveguides is as strong as expected from the exponential mode decay or whether it is perturbed by the central site. As many larger waveguide lattices contain such three-site units, the results of our study will also be applicable to extended systems.

We start with a linear chain of three identical single-mode waveguides with spacing $d$, as sketched in Fig.~\ref{Setup}\textbf{(a)}. 
In the paraxial approximation and the tight-binding regime, one can describe the system by discrete field amplitudes $\psi_{1,2,3}(z)$, whose evolution along the longitudinal coordinate $z$ is governed by:
\beq
\begin{aligned}
i\frac{\mathrm{d}\psi_{1,3}}{\mathrm{d}z}+\kappa_1\psi_2+\kappa_2\psi_{3,1}=0 \\
i\frac{\mathrm{d}\psi_2}{\mathrm{d}z}+\kappa_1\left(\psi_1+\psi_3\right)=0.
\end{aligned}
\label{tightbinding}
\eeq 
The dynamics of the system is usually dominated by the coupling between nearest neighbours $\kappa_1$ \cite{Somekh:ChannelOpticalWaveguides}. Due to the exponential decay of the mode fields away from the waveguides, the second-order coupling between the outer sites $\kappa_2$ is much weaker. Therefore, it influences the propagation dynamics only slightly, which makes $\kappa_2$ hard to measure directly in light propagation experiments. In fact, the impact of $\kappa_2$ on the light propagation is indistinguishable from the one of the central site $2$ being detuned with respect to the outer ones, a situation which can occur in laser-written waveguides due to stress fields \cite{Fernandes:BirefringenceTuningStressField,PerezLeija:StateTransfer}.
 
For acquiring experimental access to the second-order coupling, which is not obstructed by nearest-neighbour coupling dynamics, we resort to stationary solutions of the coupled system. They should have the form $\left(\psi_1,\psi_2,\psi_3\right)^{\intercal}=\mathbf{u}\mathrm{e}^{i\beta z}$ with $\mathbf{u}=\left(u_1,u_2,u_3\right)^{\intercal}$ being the $z$-independent eigenmode. Substitution into Eq.~(\ref{tightbinding}) yields the eigenvalue problem:
\beq
\left(\begin{array}{ccc}
0 & \kappa_1 & \kappa_2\\
\kappa_1 & 0 & \kappa_1\\
\kappa_2 & \kappa_1 & 0
\end{array}\right)\mathbf{u}
=\beta\mathbf{u}.
\eeq
In this notation, the eigenvalues $\beta$ are expressed relative to the propagation constant of a single waveguide. One finds rather straightforwardly that the solution bearing the second-largest eigenvalue $\beta_2$ is always:
\beq
\mathbf{u_2}\propto\left(1,0,-1\right)^{\intercal};\;\beta_2=-\kappa_2.
\label{SecondEigenmode}
\eeq
Therefore, the eigenvalue of this antisymmetric eigenmode can be used as a probe for the strength of the second-order coupling in the system. Note that Eq.~(\ref{SecondEigenmode}) holds independently of the particular value of $\kappa_1$, as long as the configuration remains symmetric with respect to the central site. Moreover, one can show that neither a detuning of the inner site nor nearest-neighbour mode overlaps, as considered in Ref.~\onlinecite{Heilmann2014}, have an impact on the antisymmetric eigenmode $\mathbf{u_2}$ and its eigenvalue.
As a reference system we consider a triangular configuration, as shown in Fig.~\ref{Setup}\textbf{(b)}, with the same distance $2d$ between the outer sites but some vertical offset $b$ of the central channel. If the inner waveguide has any perturbative influence on $\kappa_2$, one can intuitively expect it to be weaker in the triangle than in the linear array, due to an increased nearest-neighbour distance and the unobscured \lq line of sight\rq\ between the outer sites.

\bfig
\includegraphics[width=\columnwidth]{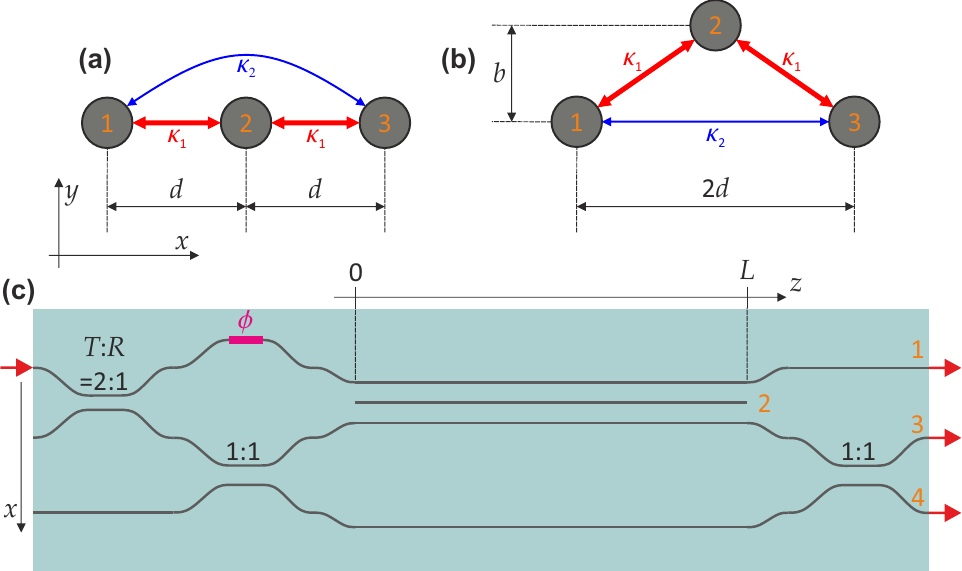}
\caption{\label{Setup}System under investigation. \textbf{(a)} Cross-section of three waveguides arranged in a linear chain. First- and second-order coupling are indicated by red and blue arrows, respectively. \textbf{(b)} Cross-section of a triangular configuration with height $b$. \textbf{(c)} Top-view of the interferometer probing the eigenvalue of the second eigenmode of the three waveguides. The couplers are specified with their ratio of transmitted ($T$) vs. reflected ($R$) power. All waveguides have constant and equal refractive index, except at the phase shifter (magenta section).}
\efig
In order to measure the eigenvalue $\beta_2$ experimentally, the coupled three-waveguide system of length $L$ is embedded into an interferometer which excites the antisymmetric eigenmode, such that a phase $\beta_2L$ is accumulated during the propagation. If the output light is superposed with a phase reference, one can measure that phase, and, thereby, obtain the eigenvalue and the second-order coupling. Here we implement an integrated version of such an interferometer. Its layout is shown in Fig.~\ref{Setup}\textbf{(c)}. An additional waveguide channel $4$ serves as phase reference, whereas the central channel $2$ does not reach either end of the chip. The two directional couplers before $z=0$ are used to distribute the light evenly across the two sites occupied by mode $\mathbf{u_2}$ and the phase reference. If light is injected into the first waveguide, the light amplitude at $z=0$ reads $\mathbf{E}\left(0\right)\propto\left(\mathrm{e}^{i\phi},0,i,-1\right)^{\intercal}$, taking into account the intrinsic phase shift of each coupler \cite{Okamoto:Fundamentals}. The additional phase shift $\phi$ is introduced to match the eigenmode $\mathbf{u_2}$ by setting $\phi=-\pi/2$. If this phase is set correctly, the field does not change its shape while propagating through the coupled structure, but merely acquires a phase $\beta_2L$: $\mathbf{E}\left(L\right)\propto\left(-i\mathrm{e}^{i\beta_2L},0,i\mathrm{e}^{i\beta_2L},-1\right)^{\intercal}$. Thus, after the final coupler the output state of the interferometer reads $\mathbf{E}_{\mathrm{out}}\propto\left[\sqrt{2}\mathrm{e}^{i\beta_2L},0,1-\mathrm{e}^{i\beta_2L},-i\left(1+\mathrm{e}^{i\beta_2L}\right)\right]^{\intercal}$, which yields for the ratio of output powers in channels $3$ and $4$:
\beq
\frac{P_3}{P_4}=\left|\frac{E_{\mathrm{out},3}}{E_{\mathrm{out},4}}\right|^2=\tan^2\left(\frac{\beta_2}{2}L\right),
\label{powerratio}
\eeq
such that the magnitude of $\beta_2$ can be determined from that ratio.

We have implemented this eigenmode interferometer by means of direct laser waveguide writing \cite{Miura:LaserWrittenWaveguides}. The waveguides were inscribed in fused silica (Corning 7980 Standard, bulk refractive index $n_0=1.458$) by translating the material with $\unit[2.5]{mm/s}$ through the focus (numerical aperture $0.35$) of $\unit[180]{fs}$ long $\unit[800]{nm}$ laser pulses with a repetition rate of $\unit[100]{kHz}$ and average power of $\unit[25]{mW}$. The probe light had a wavelength of $\unit[633]{nm}$ and was linearly polarised along the $x$-axis in Fig.~\ref{Setup}. The length of the inner part of the interferometer was $L=\unit[7.17]{cm}$, where adjacent waveguides were spaced by $d=\unit[16]{\textrm{\textmu}m}$. The light evolution in the sample was measured via the fluorescence of colour centres \cite{Dreisow:DynamicLocalization} and the output light was imaged directly onto a camera. With these techniques at hand, the geometries of the couplers were optimised beforehand to achieve the desired power splitting ratios of $2:1$ and $1:1$. The phase $\phi$ was adjusted by writing a section of $\unit[1]{cm}$ length with an increased speed, thus reducing the waveguide's propagation constant locally and inducing a negative phase shift. A local velocity of $\unit[3.5]{mm/s}$ was found to yield the cleanest excitation of the antisymmetric eigenmode $\mathbf{u_2}$. 

\begin{figure*}
\includegraphics[width=\textwidth]{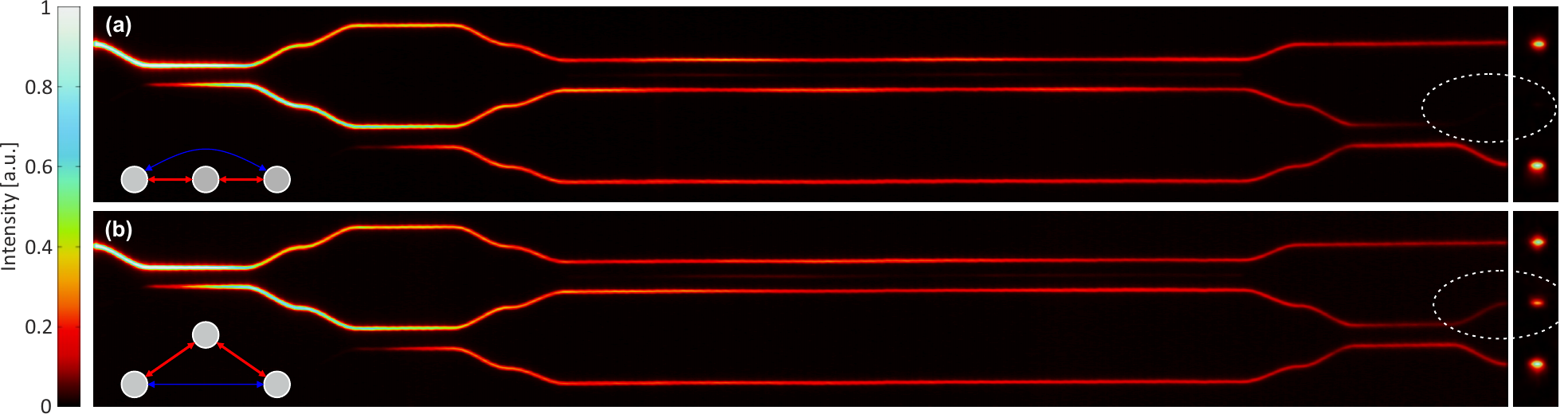}
\caption{\label{Observation}Measured light intensity evolution in the interferometer for \textbf{(a)} the linear array ($b=0$) and \textbf{(b)} the triangular configuration ($b=\unit[12]{\textrm{\textmu}m}$). The respective output intensities are shown on the right. The white ellipses highlight the key difference between the two cases in the occupation of the third mode at the output of the device. Each image has been rescaled to its maximum and the propagation images have been corrected for propagation losses.}
\end{figure*}
The experimental observations for the linear as well as for a triangular waveguide configuration are presented in Fig.~\ref{Observation}. In both cases, the initial couplers distribute the light with approximately equal power into the three channels $1$, $3$ and $4$. The antisymmetric eigenstate is predominantly excited, as indicated by the near absence of fluctuations of the light distribution in the coupled region.
For the linear chain (see Fig.~\ref{Observation}\textbf{(a)}), one observes that the output port $3$ is nearly empty, whereas it is occupied by a considerable amount of light in case of the triangular coupled system (\textbf{(b)}). This suggests that there must be a substantial difference in second-order coupling between the two cases.

\bfig[h]
\includegraphics[width=\columnwidth]{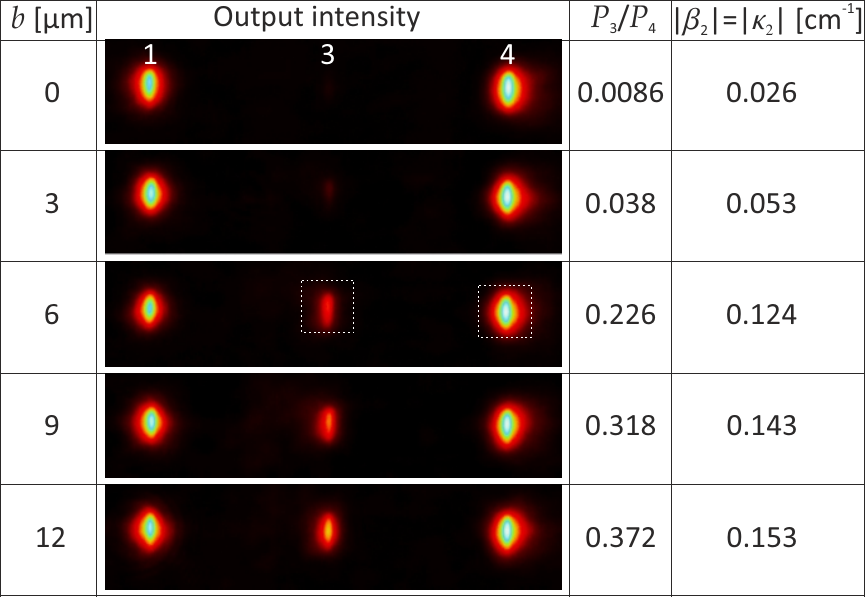}
\caption{\label{Output} Output intensity and analysis. The table shows the measured output intensities for the linear chain (top row) and triangles of increasing height $b$. The white boxes indicate the windows of integration ($51 \times 51$ pixels around the intensity maxima) which have been used to extract the output powers of all waveguides. The same colormap as in Fig.~\ref{Observation} is used. The values for $\kappa_2$ shown in the last column were obtained from the extracted output powers via inversion of Eq.~(\ref{powerratio}).}
\efig
For a quantitative analysis, output images from a range of structures with increasing height of the triangle ($b=\unit[0,3,\ldots,12]{\textrm{\textmu}m}$) were recorded. The total power guided in each channel was inferred from an integration of the measured intensity over the waveguide after correcting for background light and noise filtering. The raw images as well as the results of their analysis are listed in Fig.~\ref{Output}. One observes a clear trend towards stronger second-order coupling for increasing values of $b$, even though the separation between the outer waveguides remains the same. In the chain, $\kappa_2$ is reduced by a factor of $6$ compared to the highest triangle with $b=\unit[12]{\textrm{\textmu}m}$, which implies that the presence of the central channel has an inhibitive effect on the second-order coupling in the investigated system. Note that even though the sign of $\kappa_2$ cannot be resolved by the interferometer, a positive coupling $\kappa_2>0$ seems most plausible in all configurations, as negative coupling between waveguides has so far been demonstrated to exist only under very special circumstances \cite{Longhi:DynLoc,EfremidisAlternatingCoupling,Zeuner:NegCoupDefects}, which are not met here. Due to a high signal-to-noise ratio on the camera, imprecisions of $\kappa_2$ originating from the light intensity measurements are negligibly small. Instead, the main experimental uncertainties in our system arise from fabrication tolerances: The precision of the transverse waveguide positions is $\unit[0.3]{\textrm{\textmu}m}$, whereas each waveguide's propagation constant is uncertain by $\approx\unit[0.02]{cm^{-1}}$. These tolerances lead to some random deviations from the design structure, namely a variability in the splitting ratios of the directional couplers and additional phase shifts in the interferometer. Their impact on the experimental results is taken into account via numerical simulations of the light propagation through ensembles of devices with random realisations of these imperfections and the measured $\kappa_2$ values as model input. As most imperfections tend to decrease the interference contrast, one obtains a small systematic shift in the resulting $\kappa_2$, which can be subtracted from the measured data for bias-correction. The corrected data and their uncertainties (obtained from the standard deviations of the ensembles) are shown in Fig.~\ref{Simulation}\textbf{(a)}, together with the measured raw values. For $b=0$ and $b=\unit[3]{\textrm{\textmu}m}$, the observed inhibition of second-order coupling is clearly significant. 

Finally, we compare the experimental observations with the amount of coupling expected in absence of the central channel, i.e., in a directional coupler with gap $2d$. In order to calculate this coupling rate, we reconstruct the refractive index profile of a single waveguide from its measured mode intensity distribution. This reconstruction is performed via inversion of the Helmholtz equation \cite{Mansour:RefractiveIndexMeasurement}, again after appropriate noise filtering of the image. One obtains an elliptic profile with a maximum height of about $7\times 10^{-4}$ and a full-width-half-maximum area of $\unit[3\times 11.5]{\textrm{\textmu}m}$, as shown in Fig.~\ref{Simulation}\textbf{(b)}. A system of two such waveguides features a symmetric and an antisymmetric eigenmode with eigenvalues $\beta_{\mathrm{s}}$ and $\beta_{\mathrm{as}}$, respectively. Their eigenvalue difference gives the coupling rate \cite{Okamoto:Fundamentals}:
\beq
\kappa_{\mathrm{dc}}=\frac{\beta_{\mathrm{s}}-\beta_{\mathrm{as}}}{2}.
\label{CouplerModeSplitting}
\eeq
We calculate the modes numerically from the reconstructed index profile by solving the full vectorial Helmholtz equation via a finite element method (COMSOL). The two eigenmodes polarised in $x$-direction and their eigenvalues are plotted in Fig.~\ref{Simulation}\textbf{(c)}. One obtains via Eq.~(\ref{CouplerModeSplitting}) $\kappa_{\mathrm{dc}}\approx \unit[0.126]{cm^{-1}}$. This is the second-order coupling one would expect in our three-site system if it was purely governed by the exponential mode decay of the outer modes and the central waveguide had no influence at all. Its value is indicated by the horizontal line in Fig.~\ref{Simulation}\textbf{(a)}. For triangles with sufficient vertical displacement ($b\geq \unit[6]{\textrm{\textmu}m}$), the measured second-order couplings are in accordance with this non-pertubation scenario. However, when the central site is placed between the outer channels, its presence suppresses the second-order coupling.
\bfig
\includegraphics[width=\columnwidth]{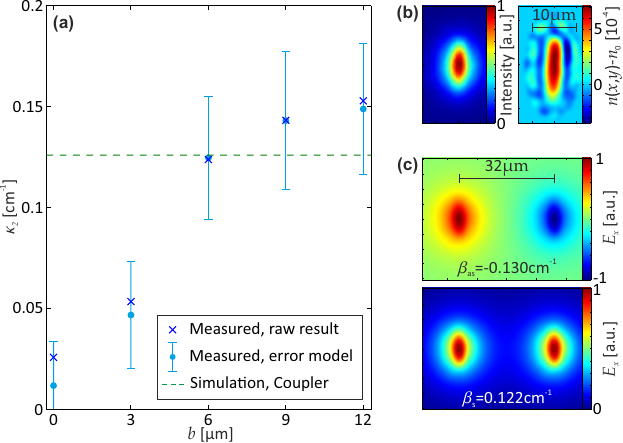}
\caption{\label{Simulation} \textbf{(a)} Comparison of the raw experimental data (blue crosses; from last column of Fig.~\ref{Output}), the experimental data after error analysis (cyan circles and error bars) and the calculated coupling rate of a two-site coupler (green line). The error-modelled data points are shifted from the raw data according to the biases obtained from error simulation while the error bars indicate $\pm 1\sigma$ of the random uncertainty. \textbf{(b)} Measured mode intensity profile of waveguide channel $1$ and reconstructed refractive index profile. \textbf{(c)} Calculated eigenmodes and eigenvalues of a coupler with $2d=\unit[32]{\textrm{\textmu}m}$.}
\efig

In this work, we have presented an integrated eigenmode interferometer, which permits to measure eigenvalues of coupled waveguide systems directly. In particular, it is used here for determining the second-order coupling in a system of three coupled channels. It has been shown that the second-order coupling equates to the magnitude of one of the system's eigenvalues. This eigenvalue is independent of the first-order coupling, therefore the second-order coupling is not hidden by other effects as it is often the case in propagation experiments. We found that in a laser-written linear chain of three waveguides the second-order coupling is strongly inhibited. The physical mechanism behind this inhibition remains elusive at this point. Numerical simulations of the three-site system suggest a strong dependence of the inhibition on the precise form of the refractive index profiles. Therefore, an extended experimental study with different fabrication parameters as well as analogue investigations in different physical systems, such as lithographic arrays \cite{Pertsch:BlochOscillationTemp,Morandotti:BlochOscillation,Longhi:DynLoc}, fiber waveguides \cite{Roepke:FiberWaveguideArray} or microwave resonators \cite{Bellec:MicrowaveTightBinding}, could provide more insight in this respect. 
A suppression of coupling by a buffer structure could perhaps be exploited to reduce undesired cross-talk at waveguide array junctions \cite{Keil:RoutingSwitching} or waveguide crossings \cite{Meany:3DMulitportQuantumInterference,Jones:TwoLayerWaveguideCrossing,Crespi:FourierSuppressionLaw} within three-dimensional photonic routing networks. 
From a more general perspective, the concept of the interferometer itself is applicable to any eigenmode of any tight-binding coupled waveguide system: Given a certain coupling configuration, only the first part of the interferometer has to be engineered to excite the desired eigenmode. Moreover, reconfigurable optical circuits \cite{Chaboyer:TunableThreePath,Carolan:UniversalLinearOptics} could be employed to allow excitation and measurement of all eigenmodes in a single device. 

The authors thank Marco Ornigotti for valuable discussions and acknowledge financial support by the European Research Council (ERC, project 257531-EnSeNa), the Canadian Institute for Advanced Research (CIFAR, Quantum Information Science Program), the German Ministry of Education and Research (Center for Innovation Competence program, grant 03Z1HN31) and the German Research Foundation (DFG, project SZ276/7-1). R.K. is supported via a Lise-Meitner-Fellowship of the Austrian Science Fund (FWF, project M 1849).
%
%



\end{document}